\documentclass[final]{statsoc}

\usepackage{epsfig}
\usepackage{bm}
\usepackage[psamsfonts]{amssymb}
\usepackage[cmex10]{amsmath}
\usepackage{algorithm}
\usepackage{setspace}
\usepackage{subfigure}

\title[Bayesian changepoint analysis for atomic force microscopy and soft material indentation]{Bayesian changepoint analysis for atomic force \\microscopy and soft material indentation}

\author[D.~Rudoy, S.~G.~Yuen, R.~D.~Howe and P.~J.~Wolfe]{\vspace{-1.5\baselineskip}%
Daniel Rudoy, Shelten G. Yuen, Robert D. Howe and Patrick J. Wolfe}
\coaddress{Patrick J. Wolfe, Statistics and Information Sciences Laboratory, Harvard University, Oxford Street, Cambridge, MA 02138-2901, USA}
\email{wolfe@stat.harvard.edu; software URL: http://sisl.seas.harvard.edu/BayesCP.html}
\address{Harvard University, Cambridge, USA.\\\emph{[Manuscript submitted for publication September 2008.  Revised September 2009.]}}
\begin{document}

\maketitle

\begin{abstract}

Material indentation studies, in which a probe is brought into controlled physical contact with an experimental sample, have long been a primary means by which scientists characterize the mechanical properties of materials.  More recently, the advent of atomic force microscopy, which operates on the same fundamental principle, has in turn revolutionized the nanoscale analysis of soft biomaterials such as cells and tissues.  This paper addresses the inferential problems associated with material indentation and atomic force microscopy, through a framework for the changepoint analysis of pre- and post-contact data that is applicable to experiments across a variety of physical scales.  A hierarchical Bayesian model is proposed to account for experimentally observed changepoint smoothness constraints and measurement error variability, with efficient Monte Carlo methods developed and employed to realize inference via posterior sampling for parameters such as Young's modulus, a key quantifier of material stiffness.  These results are the first to provide the materials science community with rigorous inference procedures and uncertainty quantification, via optimized and fully automated high-throughput algorithms, implemented as the publicly available software package BayesCP.  To demonstrate the consistent accuracy and wide applicability of this approach, results are shown for a variety of data sets from both macro- and micro-materials experiments---including silicone, neurons, and red blood cells---conducted by the authors and others. 


\end{abstract}

\section{Introduction}
\label{sec:intro}

This article develops a hierarchical Bayesian approach for contact point determination in material indentation studies and atomic force microscopy (AFM).  Contemporary applications in materials science and biomechanics range from analyzing the response of novel nanomaterials to deformation~\citep{wong97}, to characterizing disease through mechanical properties of cells, tissues, and organs~\citep{costaAFM}.  Experimental procedures and analyses, however, remain broadly similar across these different material types and physical scales~\citep{gouldstone06}, with the scientific aim in all cases being to characterize how a given material sample deforms in response to the application of an external force.

As illustrated in Figure~\ref{fig:indentCartoon} overleaf,
\begin{figure}[!t]
 \centering
 \makebox{\includegraphics[width=0.22\columnwidth, angle=90]{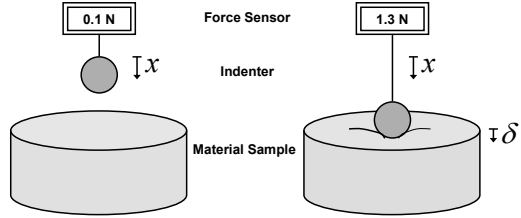}}
 \caption{\label{fig:indentCartoon}Diagram of a macro-scale indentation experiment in which a spherical probe, attached to a force sensor, indents a material sample and deforms it by a distance $\delta$. In this hypothetical example, a net change of $1.2$ Newtons in resistive force is consequently observed.}
 \vspace{-1.0\baselineskip}%
\end{figure}
indentation experiments employ a probe (or cantilever arm, in the case of AFM) to apply a controlled force to the material sample.  This indenting probe displaces the sample while concurrently measuring resistive force, with the resultant force-position data used to infer material properties such as stiffness (in analogy to compressing a spring in order to experimentally determine its spring constant).  Prior to subsequent data analysis, a key technical problem is to determine precisely the moment at which the probe comes into contact with the material.  Sample preparation techniques and sizes frequently preclude the direct measurement of this contact point, and hence its inference from indenter force-position data forms the subject of this article.

At present, practitioners across fields lack an agreed-upon standard for contact point determination; a variety of \emph{ad hoc} data pre-processing methods are used, including even simple visual inspection~\citep{Lin2007a}.  Nevertheless, it is well recognized that precise contact point determination is necessary to accurately infer material properties in AFM indentation experiments~\citep{crick07}.  For example, \citet{Dimitriadis} show that for small displacements of thin films, estimation errors on the order of $5$~nm for a $2.7$~$\mu$m sample can cause an increase of nearly $200$\% in the estimated Young's modulus---the principal quantifier of material stiffness.  When soft materials such as cells are studied at microscopic scales, for example to determine biomechanical disease markers~\citep{costaAFM}, the need for robust and repeatable AFM analyses becomes even greater~\citep{Lin2007a}.

In this article, we present the first formulation of the contact point determination task as a \emph{statistical changepoint} problem,  and subsequently employ a switching regressions model to infer Young's modulus.  Section~\ref{sec:data} summarizes the basic principles of material indentation, showing that the resultant force-displacement curves are often well described by low-order polynomials. Section~\ref{sec:Models} introduces a corresponding family of Bayesian models designed to address a wide range of experimental conditions, with specialized Markov chain Monte Carlo samplers for inference developed in Section~\ref{sec:inference}.  Following validation of the proposed inference procedures in Section~\ref{sec:validation}, they are employed in Section~\ref{sec:indentExp} to infer material properties of mouse neurons and human red blood cells from AFM force-position data.  The article concludes in Section~\ref{sec:discussion} with a discussion of promising methodological and practical extensions.

\section{Material indentation}
\label{sec:data}

\subsection{Indentation experiments and data}

\begin{figure}[!t]
  \centering\hspace{-1.6em}%
  \subfigure[\label{fig:siliconeExampleFD} Silicone indentation data]{
  \makebox{\includegraphics[width=.45\columnwidth]{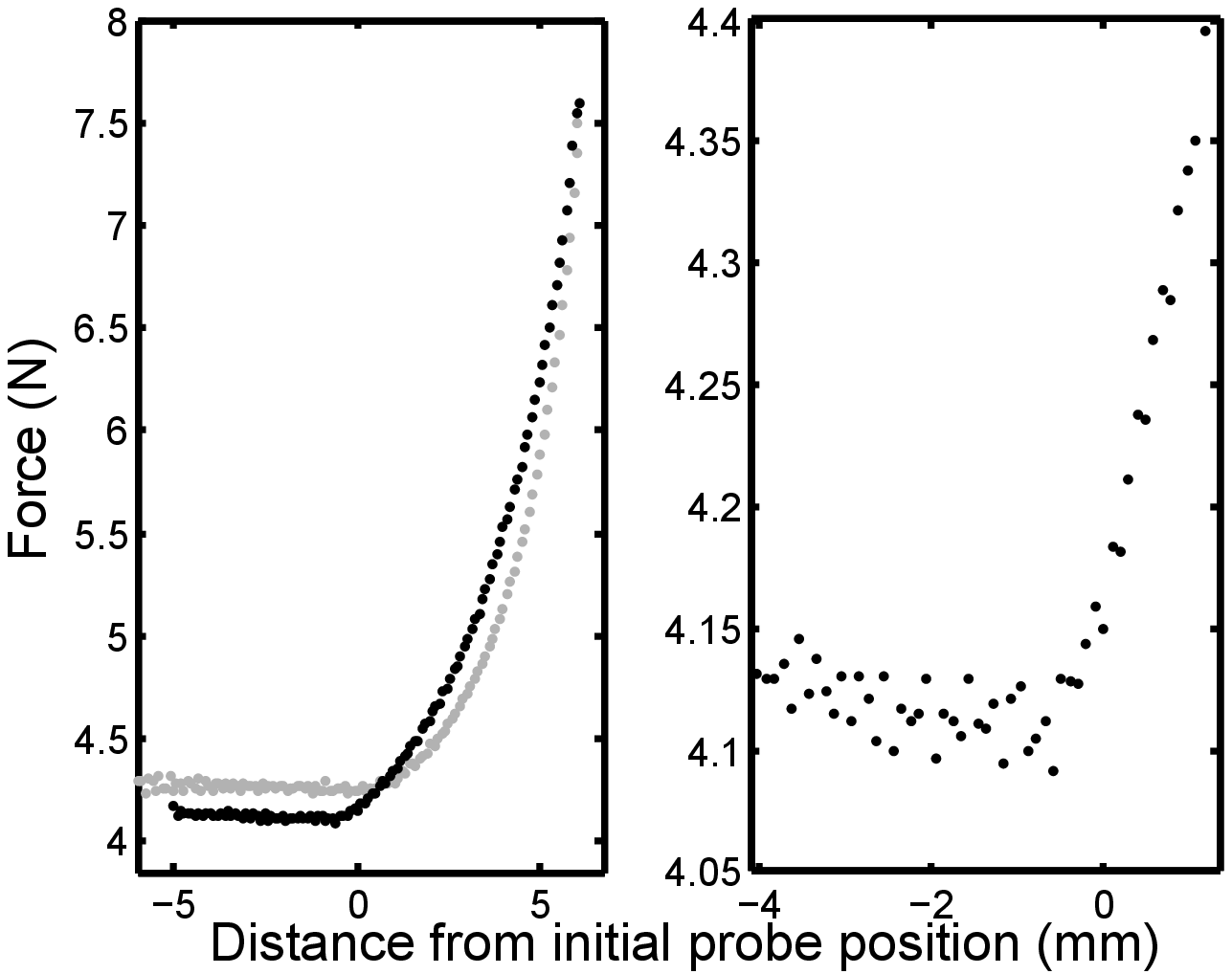}}
  }\hspace{1em}%
  \subfigure[\label{fig:rbcExampleFD} Red blood cell indentation data]{
  \makebox{\includegraphics[width=.455\columnwidth]{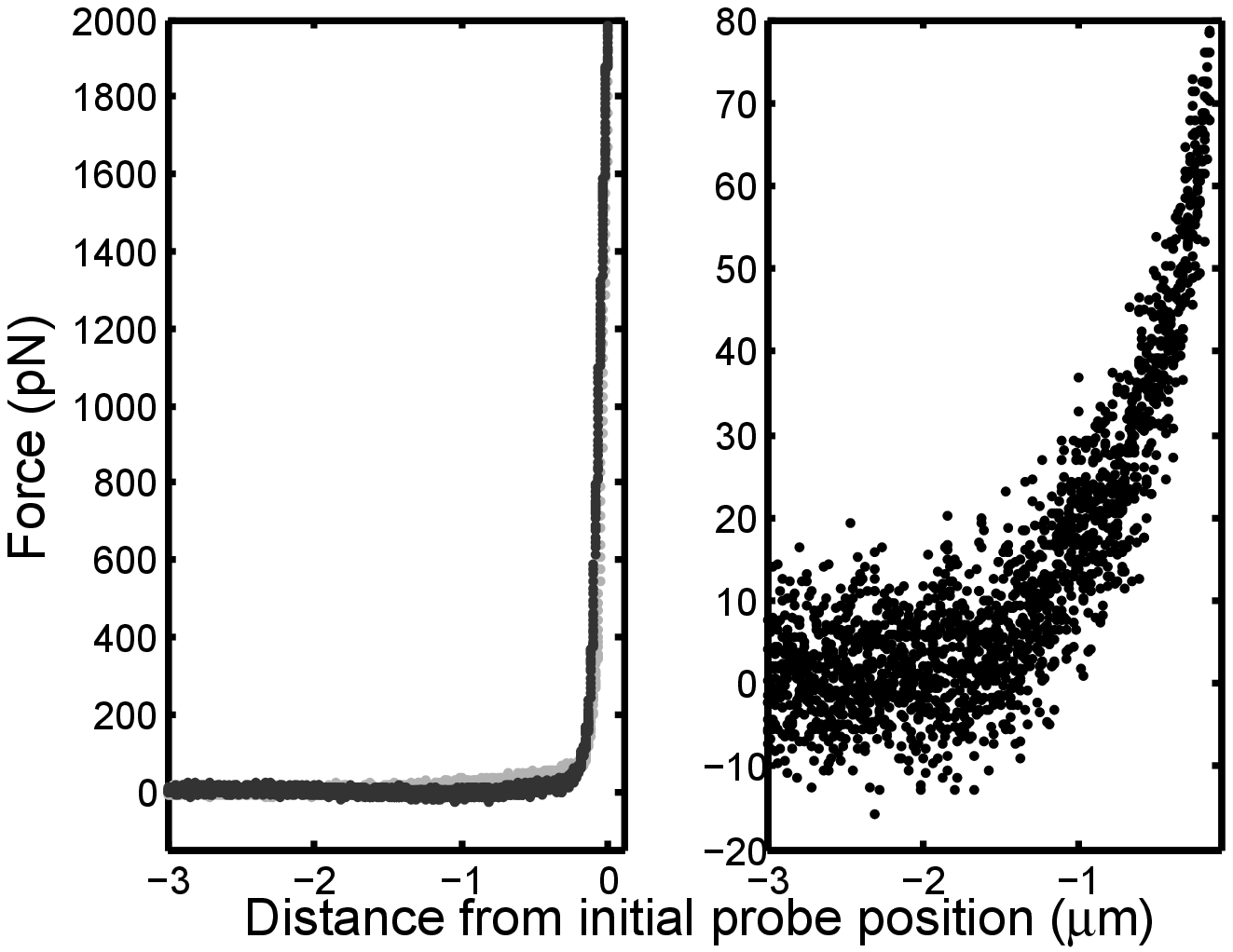}}
  }
  \caption{\label{fig:fdExample} Examples of force-position curves from which material stiffness properties are to be inferred, with force measurements during indentation and subsequent probe retraction shown in black and grey, respectively, and subsets of the indentation data shown near the presumed contact points.  Experiments were performed using a mechanical arm for the soft silicone sample, and an atomic force microscope for the red blood cell sample; note the differences in physical scale and noise level.}
  \vspace{-0.5\baselineskip}%
\end{figure}

Indentation experiments proceed by carefully moving a probe from an initial noncontact position into a material sample, as shown in Figure~\ref{fig:indentCartoon}, while measuring the resistive force at some prescribed temporal sampling rate. After a small deformation is made, the probe retracts to its initial position; during this stage the resistive force decreases with every subsequent measurement. At the conclusion of each such experiment, two force-position curves are produced, examples of which are shown in Figure~\ref{fig:fdExample}.  In this article we consider only the forward-indentation curves, as is standard practice~\citep{Lin2007a}, though the methods we present are extensible to retraction data whenever suitable models are available.%

Despite significant differences in physical scale and noise level, the curves of Figure~\ref{fig:fdExample} (and indeed, most indentation data sets) feature several common characteristics. In the pre-contact region, force response appears linear in the position of the indenter; drift due to experimental conditions is often present, yielding a non-zero slope as in Figure~\ref{fig:fdExample}(a). The post-contact data appear well modeled by low-order polynomial functions of the corresponding displacement, and indeed, the key assumption of such experiments is that, conditioned upon knowledge of the geometry of the indenting probe, the relationship between the degree of material deformation and the measured resistive force depends \emph{in a known way} on the material stiffness. This is analogous to the case in which an ideal spring is compressed a specified distance $\delta$ by a force of known magnitude; a measure of the spring's stiffness is given by the spring constant $k$, which may be calculated using Hooke's law: $F = -k\delta$.

\subsection{Contact mechanics and the Hertzian model}
\label{sec:ContactPointEst}

Indentation data such as those shown in Figure~\ref{fig:fdExample} are typically acquired for \emph{linear-elastic} materials.  This implies not only that the material instantaneously returns to its original shape following the cessation of an external force, but also that the relationship between the applied stress (force per unit area) and the resultant strain (deformation per unit length) is linear.  This ratio is known as \emph{Young's modulus}---the primary quantifier of material stiffness introduced above---and is reported in units of Pascals.

In small-deformation experiments to infer Young's modulus $E$ for linear-elastic materials, both the indenter geometry and the measured resistive force come into play, by way of the so-called \emph{Hertzian model}~\citep{Lin2007a}.  Specifically, the relationship between the sample deformation depth $\delta$ and the measured resistive force $F$ takes the form
\begin{equation}
    \label{eq:HertzModel}
    F \propto E \cdot \delta^\beta \text{,}
\end{equation}
where the constant of proportionality and the fixed parameter $\beta$ depend on the geometry of the indenter tip in a known way.  Examples include indentation by a sharp pyramid with tip angle $2\phi$, or a sphere of radius $R$, whereupon~\eqref{eq:HertzModel} takes the following forms:
\begin{equation}
    \label{eq:indenterEqns}
    F = \frac{1.5\tan\phi}{2(1-\nu^2)} \, E \cdot \delta^2
    \quad \text{(pyramid)} \qquad \text{or} \qquad
    F = \frac{4R^{1/2}}{3(1-\nu^2)} \, E \cdot \delta^{3/2}
    \quad \text{(sphere)}
    \text{,}
\end{equation}
with $\nu$ a known, dimensionless quantity termed Poisson's ratio.  A subsequent fitting of the Hertzian model of~\eqref{eq:HertzModel} to experimental data thus allows one to obtain an estimate for Young's modulus $E$ once the post-contact region has been identified.  (Below we retain the standard practitioner notation $(E,F,R,\delta,\nu,\phi)$, as distinct from other variables to follow.)

\subsection{Pre- and post-contact data regimes}
\label{sec:prePostRegimes}

Observe that the Hertzian model of~\eqref{eq:HertzModel} posits a relationship between force and indentation \emph{depth}, whereas the data of Figure~\ref{fig:fdExample} are seen to be a function of the \emph{position} of the indenter.  The Hertzian model thus describes the underlying physics of the \emph{post}-contact stage of a typical indentation experiment, while the measured data also comprise a \emph{pre}-contact stage.  As we detail below, the union of these two regimes is well described by a switching regressions scenario, with force measurements prior to contact typically linear in the position of the probe.  The corresponding intercept represents the equilibrium tare required to achieve zero net force, while the slope captures constant-velocity drift in force measurements that can arise in a variety of experimental settings.

In both the pre- and post-contact data regimes, it is standard to assume independence of measurement errors---a reasonable assumption for all practically achievable force sampling rates.  Errors prior to contact arise due to force sensor vibration in the experimental medium, thermal variations, and other effects---whereas after contact they also depend on interactions between the probe and the sample such as frictional forces.  Consequently, while error variances in these regimes can be expected to differ in practice (analysis of AFM data from a red blood cell, shown in Figure~\ref{fig:rbc3Data} and described later, reveals one such example), their relative magnitudes are not known a priori.  As a final consideration, uncertainty in the reported position of the indenting probe is typically several orders of magnitude smaller than the distance between consecutive force sampling points, and can safely be disregarded.

As noted by \citet{crick07}, all of the sources of variability mentioned above can lead to large relative errors in resistive force measurements near the contact point; Figure~\ref{fig:rbcExampleFD} illustrates a typical scenario.  This makes manual identification of the contact point difficult in many cases, and motivates the model-based approach that we now describe.

\section{Bayesian changepoint model}
\label{sec:Models}

Having outlined the basic principles of material indentation, we now proceed to formulate a Bayesian model for force-position data that encompasses both pre- and post-contact regimes.  In light of the discussion above, the corresponding task of contact point determination is recognizable as a changepoint estimation problem in the context of switching regressions.  In turn, the Hertzian model of~\eqref{eq:HertzModel} implies that an estimate of Young's modulus can be obtained as a \emph{linear} function of the leading post-contact regression coefficient.  Control over experimental conditions implies that each indentation data set contains precisely one changepoint, thus obviating any need to estimate the presence or number of contact points.

During an indentation experiment, the indenter moves continuously through a sequence of $n$ positions $\bm{x} = (x_1, x_2, \ldots, x_n)'$ and records a force measurement $y_i$ at each position $x_i$, resulting in a sequence of force measurements $\bm{y} = (y_1, y_2, \ldots, y_n)'$.  As we shall later treat models for soft material indentation, in which the pre- and post-contact curves are constrained to be continuous at the regression changepoint, we begin by introducing a continuous parameter $\gamma \in (1, n)$ denoting the \emph{contact point index}, with the corresponding  \emph{contact point} at which the indenter first contacts the sample denoted by $x_{\gamma} \in (x_1, x_n)$.

\subsection{Data likelihood for indentation experiments}
\label{sec:likelihood}

We adopt a classical switching regressions scenario for our model, in which $y$ is assumed to be a polynomial function of known degree $d_1$ in position $x$ prior to contact, and of known degree $d_2$ in deformation depth $\delta = x-x_\gamma$ after contact with the sample is made.   This formulation encompasses the Hertzian model of~\eqref{eq:HertzModel} if fractional powers are allowed; however, for clarity of presentation we consider $d_2$ to be an integer unless otherwise noted.  Letting $p = d_1+d_2+2$ denote the number of regression coefficients in our model, and with $n$ the number of data points, the corresponding design matrix is hence of dimension $n \times p$.  We subsequently employ the subscript {\scriptsize $\gamma$} to denote any quantity that depends on $\gamma$, and index via subscripts {\scriptsize $1$} and {\scriptsize $2$} the pre- and post-contact regression regimes.

We denote the regression coefficients by $\bm{\beta}_1 \in \mathbb{R}^{d_1+1}$ and $\bm{\beta}_2 \in \mathbb{R}^{d_2+1}$, with design matrices $\bm{X}_{1,\gamma}, \bm{X}_{2,\gamma}$ defined as follows, for $\lfloor \gamma \rfloor$ the largest integer less than or equal to $\gamma$:
\begin{equation}
    \label{eq:designMatrices}
    \bm{X}_{1,\gamma} =
    \begin{pmatrix} 1 & x_1 & \cdots & x_1^{d_1} \\ 1 & x_2 & \ldots & x_2^{d_1} \\
    \vdots & \vdots & \ddots & \vdots \\
    1 & x_{\lfloor \gamma \rfloor} & \cdots & x_{\lfloor \gamma \rfloor}^{d_1}
    \end{pmatrix}\!,\quad 
    \bm{X}_{2,\gamma} = \begin{pmatrix} 1 & x_{\lfloor \gamma \rfloor +1}- x_\gamma & \cdots & \left (x_{\lfloor \gamma \rfloor + 1} -x_\gamma \right )^{d_2} \\
     1 & x_{\lfloor \gamma \rfloor + 2}- x_\gamma & \ldots & \left (x_{\lfloor \gamma \rfloor + 2}- x_\gamma \right)^{d_2} \\
    \vdots & \vdots & \ddots & \vdots \\
     1 & x_n- x_\gamma & \ldots & \left (x_n-x_\gamma \right)^{d_2}
    \end{pmatrix}\! \text{.}
\end{equation}
The observed data $\bm{y}$ may likewise be partitioned into pre- and post-contact vectors
\begin{equation*}
    \bm{y}_{1,\gamma} = (y_{1}, y_2, \ldots, y_{\lfloor \gamma \rfloor})^{'} \quad \text{and} \quad
    \bm{y}_{2,\gamma} = (y_{\lfloor \gamma \rfloor+1}, y_{\lfloor \gamma \rfloor+2}, \ldots, y_{n})^{'} \text{,}
\end{equation*}
and, following our discussion in Section~\ref{sec:prePostRegimes} regarding the noise characteristics typical of indentation experiments, we assume independent and Normally distributed additive errors, with unknown variances $\sigma_1^2, \sigma_2^2$.  Thus for $1\leq i \leq n$, we have that $y_i$ is distributed as follows:
\begin{equation}
    \label{eq:dataLikelihood}
    y_{i} \sim
        \begin{cases}
            \mathcal{N}(\bm{X}_{1,\gamma}\bm{\beta}_1, \sigma_1^2) & \text{if $1 \leq i \leq \lfloor \gamma \rfloor$,} \\
            \mathcal{N}(\bm{X}_{2,\gamma}\bm{\beta}_2, \sigma_2^2) & \text{if $\lfloor \gamma \rfloor + 1 \leq i \leq n$.}
        \end{cases}
\end{equation}

Note that the statistical model of~\eqref{eq:dataLikelihood} is consistent with the Hertzian mechanics model of~\eqref{eq:HertzModel}, resulting in a post-contact force-response curve that is a power of the \emph{displacement} $\delta = x - x_\gamma$ of the material sample, rather than the \emph{position} $x$ of the indenter.  However, when $d_2$ is an integer, the coefficients of these two polynomials are related by a simple linear transformation. Consider, for instance, a quadratic curve in $\delta$ given by $f(\delta) = a_0 + a_1\delta + a_2\delta^2$.  One may rewrite $f(\delta)$ as a quadratic polynomial in $x$ as follows:
\begin{equation}
    \label{eq:coordTrans}
    a_0 + a_1(x-x_\gamma) + a_2(x-x_\gamma)^2 = b_0 + b_1x + b_2x^2 \text{,}
\end{equation}
where $b_2 = a_2$, $b_1 = a_1 - a_2x_\gamma$, and $b_0 = a_0 - a_1x_\gamma + a_2x_\gamma^2$. This transformation enables $\bm{X}_{2,\gamma}$ to be reformulated directly in terms of indenter position $x$, such that it no longer depends \emph{continuously} on $x_\gamma$, in contrast to~\eqref{eq:designMatrices}.  Transformations akin to~\eqref{eq:coordTrans} do not apply, however, when the $d_2$ is a fraction, as in the case of~\eqref{eq:indenterEqns} for a spherical indenter, or when a continuity constraint is enforced at the changepoint; we detail such cases below.

\subsection{A general parametric Bayesian model for material indentation}

The likelihood of~\eqref{eq:dataLikelihood}, together with the presence of genuine prior information dictated by the underlying physics of material indentation experiments, suggests a natural hierarchical Bayesian model. In contrast to the semi-conjugate approach taken by~\cite{carlin1992}, we detail below a fully conjugate model, as this allows for analytical simplifications that we have observed to be important in practice.  Integrating out nuisance parameters improves not only the mixing of the chains underlying the resultant algorithms and inferential procedures, but also their computational tractability when data sizes grow large.

We specify prior distributions for all model parameters, including the contact point index $\gamma \in (1, n)$, the pre- and post-contact regression coefficients $\bm{\beta}_{1}$ and $\bm{\beta}_{2}$, and the error variances $\sigma_{1}^{2}$ and $\sigma_{2}^{2}$. For $i \in \{1,2\}$, we then assume that $\bm{\beta}_i \sim \mathcal{N}(\bm{\mu}_i,\sigma_i^2 \bm{\Lambda}_i^{-1})$, with $\bm{\Lambda}_i$ a $(d_i +1) \times (d_i+1)$ diagonal positive definite matrix and $\bm{\mu}_i \in \mathbb{R}^{d_i+1}$.  A standard inverse-Gamma conjugate prior $\mathcal{IG}(a_0,b_0)$ is adopted for both variances $\sigma_1^2$ and $\sigma_2^2$. Finally, recalling that a single regression changepoint is always assumed within the data record, we employ a uniform prior distribution on the interval $(1,n)$ for the contact point index $\gamma$.  In certain experimental settings, whereupon the initial position of the indenter is known to be at least a certain distance from the sample, an informative prior distribution may well be available.

Because of sensitivity to hyperparameters, we follow standard practice and adopt hyperpriors for increased model robustness.  A Gamma prior is assumed on $b_0$ so that $b_0 \sim \mathcal{G}(\kappa, \eta)$, however, we determined from simulations that the posterior estimators considered were not sensitive to the prior parameters $\bm{\mu}_i$ and $\bm{\Lambda}_i$, and so did not employ an additional level of hyperprior hierarchy for the regression coefficients.
For notational convenience, we let $\bm{1}_{k}$ denote the 1-vector of dimension $k$ whose entries are all equal to one and $\bm{0}$ the zero matrix of appropriate dimension, and define the variables $\bm{y}$, $\bm{\beta}$, $\bm{\mu}$, $\bm{X}_{\gamma}$, $\bm{\Lambda}$, $\bm{\Sigma}_\gamma$, and $\bm{\Sigma}$ as follows:
\[
 \bm{y} = \begin{pmatrix} \bm{y}_{1,\gamma} \\ \bm{y}_{2,\gamma} \end{pmatrix} \in \mathbb{R}^{n \times 1} \text{,} \qquad \bm{\beta} = \begin{pmatrix} \bm{\beta}_1 \\ \bm{\beta}_2 \end{pmatrix}  \in \mathbb{R}^{p \times 1} \text{,} \qquad
 \bm{\mu} = \begin{pmatrix} \bm{\mu}_1 \\ \bm{\mu}_2 \end{pmatrix}  \in \mathbb{R}^{p \times 1} \text{;}
\]
\vspace{-0.5\baselineskip}%
\[
    \bm{X}_{\gamma} = \begin{pmatrix} \bm{X}_{1,\gamma} & \bm{0} \\ \bm{0} & \bm{X}_{2,\gamma} \end{pmatrix} \in \mathbb{R}^{n \times p} \text{,} \qquad
            \bm{\Lambda} =
        \begin{pmatrix}
            \bm{\Lambda}_1 & \bm{0} \\
            \bm{0} & \bm{\Lambda}_2 \\
        \end{pmatrix} \in \mathbb{R}^{p \times p} \text{;}
\]
\vspace{-0.5\baselineskip}%
\[
            \bm{\Sigma}_\gamma = \operatorname{diag} (
            \sigma_1^2\bm{1}_{\lfloor \gamma \rfloor},
            \sigma_2^2\bm{1}_{n-\lfloor \gamma \rfloor} ) \in \mathbb{R}^{n \times n} \text{,} \qquad
            \bm{\Sigma} = \operatorname{diag}(
            \sigma_1^2\bm{1}_{d_1+1} ,
            \sigma_2^2\bm{1}_{d_2+1} ) \in \mathbb{R}^{p \times p} \text{.}
\]
The posterior probability distribution of the model parameters $(\gamma, \bm{\beta}, \sigma^2_1,\sigma^2_2,b_0)$, conditioned on the observations $\bm{y}$ and the fixed model parameters $\bm{\psi} \triangleq (\bm{\mu}, \bm{\Lambda}, a_0, \kappa, \eta)$, is then: \small
\begin{align}
    p(\gamma, \bm{\beta}, \sigma^2_1,\sigma^2_2,b_0 | \bm{y}; \bm{\psi}) & \propto p(\bm{y}|\bm{\beta},\sigma^2_1,\sigma^2_2,\gamma) \, p(\bm{\beta}| \sigma^2_1,\sigma^2_2; \bm{\mu}, \bm{\Lambda}) \, p(\sigma^2_1 | b_0; a_0) \, p(\sigma^2_2 | b_0; a_0) \, p(b_0 ; \kappa, \eta) \, p(\gamma)  \nonumber \\
    & \propto \sigma_1^{-2(a_0-1)} e^{-b_0/\sigma_1^2} \cdot \sigma_2^{-2(a_0-1)} e^{-b_0/\sigma_2^2} \cdot b_0^{\kappa-1} e^{-b_0/\eta} \nonumber \\
     \cdot \left(\left |\bm{\Sigma}_{\gamma} \right |\left |\bm{\Sigma} \right |\right)^{-\frac{1}{2}} &\exp \left [ {\textstyle - \frac{1}{2}} \left\{ \left ( \bm{y} \!-\! \bm{X}_{\gamma}\bm{\beta} \right )'\bm{\Sigma}_{\gamma}^{-1} \left (\bm{y} \!-\! \bm{X}_{\gamma}\bm{\beta} \right ) + \left ( \bm{\beta} - \bm{\mu} \right )' \bm{\Sigma}^{-1}\bm{\Lambda} \left ( \bm{\beta} - \bm{\mu} \right ) \right\} \right ] \text{.} \label{eq:posterior}
\end{align} \normalsize

To confirm robustness, we also studied the effect of replacing the diagonal prior covariance $\bm{\Lambda}$ for the pre- and post-contact regression coefficients $\bm{\beta}_1$ and $\bm{\beta}_2$ with an appropriately adapted $g$-prior~\citep{Zellner86} such that $\bm{\beta}_i|\rho_i,\gamma \sim \mathcal{N} ( \bm{\mu}_i, \sigma_i^2 \rho^2_i (\bm{X}_{i,\gamma}' \bm{X}_{i,\gamma})^{-1} )$, with $\rho_i^2$ a scale parameter to which we ascribed a diffuse inverse-Gamma hyperprior. We observed no measurable effect of this change in priors on the resulting inference---further confirming the insensitivity of the adopted model to the prior covariance of the regression coefficients.   Moreover, efficient sampling from the conditional distribution of $\rho^2_i$ is precluded by its dependence on the contact point index $\gamma$, reducing the overall efficacy of this approach in the Markov chain Monte Carlo approaches to posterior sampling described in Section~\ref{sec:inference} below.

\subsection{Smoothness constraints at the changepoint}
\label{sec:smoothnessConstraints}

It is often the case that force-position curves are continuous at the contact point $x_\gamma$. Especially for soft materials such as the red blood cells we consider in Section~\ref{sec:indentExp}, it is to be expected that the change in the force measurement is smooth, and a continuity constraint can serve to regularize the solution in cases where many different fits will have high likelihood. Imposing smoothness constraints dates back to at least \citet{hudson66} who considered this constraint in deriving maximum likelihood estimators for switching regressions. More recently, \citet{stephens94} used it in a hierarchical Bayesian setting.

In our setting, according to the likelihood of~\eqref{eq:dataLikelihood}, a continuity constraint on the pre- and post-contact force-position curves at $x = x_\gamma$ implies that
\begin{equation}
    \label{eq:contConstraint}
    \beta_{10} + \beta_{11}x_\gamma + \cdots + \beta_{1d_1}x_{\gamma}^{d_1} =  \beta_{20} \text{,}
\end{equation}
where $\bm{\beta}_1 = (\beta_{10}, \beta_{11}, \ldots, \beta_{1d_1})'$ and $\bm{\beta}_2 = (\beta_{20}, \beta_{21}, \ldots, \beta_{2d_2})'$ denote the vectors of pre- and post-contact regression coefficients, respectively. Higher-order smoothness can also be imposed: we say that the
force-position curve is $s$ times continuously differentiable at $x_\gamma$ if the $s$th-order derivatives of the pre- and post-contact curves meet at $x_\gamma$, with~\eqref{eq:contConstraint} corresponding to the case $s = 0$. On the other hand, if $X_{2,\gamma}$ were a function of the position $x$, rather than the displacement $x-x_\gamma$, then the continuity constraint would become:
\begin{equation}
    \label{eq:contConstraint2}
    \sum_{i=0}^{d_1}\beta_{1i}x_{\gamma}^i = \sum_{j=0}^{d_2}\beta_{2j}x_{\gamma}^j \text{.}
\end{equation}
Either continuity constraint implies that the likelihood function is \emph{nonlinear} in the contact point $x_\gamma$; enforcing more degrees of smoothness at the changepoint serves to exacerbate the nonlinearity and makes the design of efficient inference methods increasingly difficult. \citet{stephens94} imposed~\eqref{eq:contConstraint2} in a Bayesian switching regressions setting and proposed a rejection sampling step within a Gibbs sampler to address the resultant nonlinearity.  Later, in Section~\ref{sec:inference}, we describe a more efficient approach  that can be applied when either~\eqref{eq:contConstraint} or~\eqref{eq:contConstraint2} (or higher-order analogues) are enforced.

\subsection{Changepoint estimation and contact point determination in the literature}
\label{sec:history}

As demonstrated above, inference for material indentation data is well matched to classical statistical frameworks for changepoint estimation.  Independent of the specifics of our contact point problem, the last half century has seen a vast body of work in this area. Sequential and fixed-sample-size varieties have been considered from both classical and Bayesian viewpoints, with numerous parametric and nonparametric approaches proposed.  We refer the interested reader to several excellent surveys, including those by~\cite{Hinkley1980},~\cite{Zacks83},~\cite{Wolfe84},~\cite{carlin1992}, and~\cite{Lai95}.  Some of the earliest work on maximum likelihood estimation of a single changepoint between two polynomial regimes was done by~\cite{Quandt58} and~\cite{Robison64}.

Historically,~\cite{chernoff64} were among the first to consider a parametric Bayesian approach to changepoint estimation. Changepoints arising specifically in linear models have been treated by many authors, including \cite{bacon71},~\cite{ferreira75},~\cite{smith75},~\cite{broemling80},~\cite{smith80}, and~\cite{menzefricke81}. The introduction of Markov chain Monte Carlo methods has led to more sophisticated hierarchical Bayesian models for changepoint problems, beginning with the semi-conjugate approach taken by~\cite{carlin1992}, in which the prior variance of the regression coefficients is left unscaled by the noise variance.  Advances in trans-dimensional simulation methods have rekindled interest in multiple changepoint problems, as discussed by~\cite{stephens94},~\cite{punskaya02} and~\cite{Fearnhead06}, among others.

In the context of material indentation, however, existing approaches to contact point determination do not make use of changepoint estimation methodology. In fact, current methods are error-prone and labor intensive---even consisting of visual inspection and manual thresholding~\citep{Lin2007a}.  However, as described earlier, the many sources of variability in indentation data imply that one cannot always simply proceed ``by eye.'' Moreover, in the context of atomic force microscopy, most experiments aiming to characterize cell stiffness, for example, employ multiple, \emph{repeated} indentations at different spatial locations. These requirements have motivated a more recent desire for effective, high-throughput \emph{automated} techniques, as detailed in \citep{Lin2007a}.

Interpreted in a statistical context, the procedures thus far adopted by practitioners fall under the general category of likelihood fitting.  \citet{rotsch99} suggest simply to take two points in the post-contact data and solve for $E$ and $\gamma$ using the appropriate Hertzian model; however, the resultant estimate of Young's modulus depends strongly on the indentation depth of the selected points~\citep{costaAFM}. \citet{costa} propose to minimize the mean-squared error of a linear pre-contact and quadratic post-contact fit to the indentation data, though under the assumption of equal pre- and post-contact variances.

None of the existing approaches adopted by practitioners, however, provides any means of quantifying uncertainty in changepoint estimation---an important consideration in practice, since measurement errors can be large relative to the reaction force of the probed material, and consequently may result in poor point estimates~\citep{crick07}. Moreover, such approaches fail to capture necessary physical constraints of the material indentation problem, such as the smoothness constraints described in Section~\ref{sec:smoothnessConstraints} above.  Such shortcomings provide strong motivation for the hierarchical model developed above, as well as the robust and automated fitting procedures we describe next.

\vspace{-.2\baselineskip}%
\section{Posterior inference via Markov chain Monte Carlo}
\vspace{-.05\baselineskip}%
\label{sec:inference}

The hierarchical Bayesian modeling framework introduced above features a large number of unknowns, with constraints on certain parameters precluding closed-form expressions for the marginal posteriors of interest.  These  considerations suggest a simulation-based approach to inference; indeed, it is by now standard to use Markov chain Monte Carlo methods to draw samples from the posterior in such cases. Though widely available software packages for Gibbs sampling are adequate for inference in certain hierarchical Bayesian settings, the complexity of the conditional distributions we obtain here (after imposing constraints and integrating out parameters whenever possible) necessitates explicit algorithmic derivations on a case-by-case basis.  To this end, we build upon the approaches of~\cite{carlin1992} and~\cite{stephens94}, and employ Metropolis-within-Gibbs techniques to draw samples from the posterior of~\eqref{eq:posterior} as well as under the smoothness constraints of Section~\ref{sec:smoothnessConstraints}.

\subsection{Metropolized Gibbs Samplers and Variance Reduction}

    The selection of conjugate priors in our model allows nuisance parameters to be integrated out, in order to reduce the variance of the resultant estimators. Following standard manipulations, we marginalize over the pre- and post-contact regression coefficients $\bm{\beta}_1$ and $\bm{\beta}_2$, respectively. This yields the following marginal posterior probability distribution: \small
    \begin{multline}
        \label{eq:RB1Posterior}
         p(\gamma, \sigma_1^2, \sigma_2^2, b_0 | \bm{y}; \bm{\psi}) \propto \sigma_1^{-2(a_0-1)} e^{-b_0/\sigma_1^2} \cdot \sigma_2^{-2(a_0-1)} e^{-b_0/\sigma_2^2} \cdot b_0^{\kappa - 1} e^{-b_0/\eta} \\
         \cdot \left(\left |\bm{\Sigma}_{\gamma} \right | \left |\bm{\Sigma} \right | \left |\bm{A}_{\gamma} \right |\right)^{-\frac{1}{2}} \exp \left \{ {\textstyle - \frac{1}{2}} \left (\bm{y}'\bm{\Sigma}_{\gamma}^{-1}\bm{y} + \bm{\mu}'\bm{\Sigma}^{-1}\bm{\mu} - {\bm{b}_{\gamma}}' \bm{A}_{\gamma}^{-1}\bm{b}_{\gamma} \right ) \right \} \text{,}
    \end{multline} \normalsize
where $\bm{A}_{\gamma} \triangleq {\bm{X}_{\gamma}}'\bm{\Sigma}_{\gamma}^{-1}\bm{X}_{\gamma} + \bm{\Sigma}^{-1}\bm{\Lambda} \in \mathbb{R}^{p \times p}$ is block-diagonal and $\bm{b}_{\gamma} \triangleq \bm{\Lambda}\bm{\Sigma}^{-1}\bm{\mu} + {\bm{X}_{\gamma}}'\bm{\Sigma}_{\gamma}^{-1}\bm{y} \in \mathbb{R}^{p \times 1}$.  The marginal posterior of~\eqref{eq:RB1Posterior} factors into a Gamma density in $b_0$, and inverse-Gamma densities in $\sigma_1^2$ and $\sigma_2^2$ by way of the following partitions of $\bm{A}_{\gamma}$ and $\bm{b}_{\gamma}$:
\begin{equation}
\label{eq:Abpart}
    \bm{A}_{\gamma} =  \begin{pmatrix}
            \bm{A}_{1,\gamma} & \bm{0} \\
            \bm{0} & \bm{A}_{2,\gamma} \\
        \end{pmatrix}, \,\,
        \begin{matrix}
        \!\!\!\!\!\!\!\!\!\!\bm{A}_{1,\gamma} \in \mathbb{R}^{(d_1+1) \times \lfloor \gamma \rfloor} \\
        \bm{A}_{2,\gamma}\in\mathbb{R}^{(d_2+1) \times (n- \lfloor \gamma \rfloor)}
        \end{matrix}\,;  \quad
        \bm{b}_\gamma =
            \begin{pmatrix}
                \bm{b}_{1,\gamma} \\
                \bm{b}_{2,\gamma}
        \end{pmatrix}, \,\,
        \begin{matrix}
        \!\!\!\!\!\!\!\!\!\!\bm{b}_{1,\gamma} \in \mathbb{R}^{\lfloor \gamma \rfloor \times 1} \\
        \bm{b}_{2,\gamma}\in\mathbb{R}^{(n- \lfloor \gamma \rfloor) \times 1}
        \end{matrix}\, \text{.}
\end{equation}
The expressions of~\eqref{eq:RB1Posterior} and~\eqref{eq:Abpart} lead to the following Gibbs sampler:
\begin{algorithm}
\caption{\label{alg1}Gibbs sampler for changepoint estimation}
\begin{enumerate}
\item Draw $\gamma \sim p(\gamma | \sigma_1^2, \sigma_2^2, b_0, \bm{y}; \bm{\psi})$ according to~\eqref{eq:RB1Posterior};
\item Draw $\sigma^{2}_1 \sim \mathcal{IG}\!\left( {a_0 + \frac{1}{2}\lfloor \gamma \rfloor, b_0 + \frac{1}{2} ({\bm{y}_{1,\gamma}}'\bm{y}_{1,\gamma} + {\bm{\mu}_1}' {\bm{\mu}_1} - {\bm{b}_{1,\gamma}}' \bm{A}_{1,\gamma}^{-1}\bm{b}_{1,\gamma} )}\right)$;
\item Draw $\sigma^{2}_2 \sim \mathcal{IG}\!\left({a_0 + \frac{1}{2}(n-\lfloor \gamma \rfloor), b_0 + \frac{1}{2} ({\bm{y}_{2,\gamma}}'\bm{y}_{2,\gamma} + {\bm{\mu}_2}' {\bm{\mu}_2} - {\bm{b}_{2,\gamma}}'\bm{A}_{2,\gamma}^{-1}\bm{b}_{2,\gamma} )}\right)$;
\item Draw $b_0 \sim \mathcal{G}\!\left(\kappa, \eta^{-1} + \sigma_1^{-2} + \sigma_2^{-2}\right)$.
\end{enumerate}%
\vspace{-0.75\baselineskip}%
\end{algorithm}

To simulate from the conditional distribution of $\gamma$, we employ as a Metropolis-within-Gibbs step a mixture of a local random walk move with an independent Metropolis step in which the proposal density is a pointwise evaluation of~\eqref{eq:RB1Posterior} on the grid $1, 2, \ldots, n$ of indenter location indices. It is also possible to integrate out both noise variances (or the hyperparameter $b_0$). In this case, additional Metropolis steps are required, as the resulting conditional density of $b_0$ is nonstandard. Our simulation studies confirm that these variants exhibit less Monte Carlo variation than a Gibbs sampler based on the full posterior of~\eqref{eq:posterior}.

\subsection{Posterior inference in the presence of smoothness constraints}
\label{sec:infSmoothness}
Bearing in mind the underlying physics of soft materials, a sufficiently high sampling rate of the force sensor relative to the speed of the indenter may yield data consistent with a smoothness assumption of a given order $s$. In this setting, our inferential procedure may be modified accordingly to appropriately take this into account. \citet{stephens94} considers continuity-constrained switching linear regressions, and uses rejection sampling to draw from the conditional distribution of the changepoint $\gamma$ given the remaining model parameters.  A more effective procedure, however, is to transform the data such that all but one of the regression coefficients may be integrated out; this variance reduction yields a Gibbs sampler analogous to Algorithm~\ref{alg1} which we detail below. In fact, it is possible to derive such an algorithm for any value of $s \geq 0$, though for ease of presentation we first describe the case $s=0$, whereupon only continuity is enforced.

This approach to variance reduction in the presence of smoothness constraints proceeds as follows: define $\widetilde{\bm{\beta}}_1 = \bm{\beta}_1$ and, without loss of generality, let $\widetilde{\bm{\beta}}_2$ contain the last $d_2$ elements of $\bm{\beta}_2$, so that $\widetilde{\bm{\beta}} = (\widetilde{\bm{\beta}}_1^{\,'}, \widetilde{\bm{\beta}}_2^{\,'})^{'}$ contains all $p = d_1+d_2+1$ independent  coefficients.  Via the continuity constraint of~\eqref{eq:contConstraint}, define a \emph{linear} transformation $\bm{T}_{\gamma}$ of $\widetilde{\bm{\beta}}$ to $\bm{\beta}$ as follows:
\begin{equation}
\label{eq:Transform}
   \bm{\beta} = \bm{T}_{\gamma}
   \widetilde{\bm{\beta}}; \qquad \bm{T}_{\gamma} \triangleq
   \begin{pmatrix}
       \bm{I}_{(d_1+1) \times (d_1 + 1)}   & \bm{0}_{(d_1 + 1) \times d_2} \\
       \bm{c}_{1\times (d_1 + 1)}         & \bm{0}_{1 \times d_2} \\
       \bm{0}_{d_2 \times (d_1 + 1)} & \bm{I}_{d_2 \times d_2}
   \end{pmatrix}\text{,}
\end{equation}
where $\bm{I}_{m \times m}$ is the $m \times m$ identity matrix, $\bm{0}_{m \times n}$ is the $m \times n$ matrix of zeros, and $\bm{c}_{1\times (d_1 + 1)} \triangleq (1, x_{\gamma}, x^2_{\gamma}, \ldots, x^{d_1}_{\gamma})$. The choice of which of the $d_1+2$ regression coefficients to select as the dependent variable is made without loss of generality, since a transformation similar to~\eqref{eq:Transform} can be defined for every such choice, as well as for the continuity constraint of~\eqref{eq:contConstraint2}.

Since one of the regression coefficients is now a \emph{deterministic} function of the others and the changepoint, we place a prior directly on $\widetilde{\bm{\beta}}$ rather than on $\bm{\beta}$. We assume $\widetilde{\bm{\beta}}_1 \sim \mathcal{N}(\bm{\mu}_1,\sigma_1^2 \bm{\Lambda}_1^{-1})$ and $\widetilde{\bm{\beta}}_2 \sim \mathcal{N}(\widetilde{\bm{\mu}}_2,\sigma_2^2 \widetilde{\bm{\Lambda}}_2^{-1})$, with $\widetilde{\bm{\Lambda}}_2$ a $d_2 \times d_2$ diagonal positive definite matrix.  Using the transformation $\bm{T}_\gamma$ of~\eqref{eq:Transform}, we obtain in analogy to~\eqref{eq:RB1Posterior} the full posterior
\begin{multline}
    \label{eq:contPosterior}
     p(\gamma, \widetilde{\bm{\beta}}, \sigma_1^2, \sigma_2^2, b_0 | \bm{y}; \bm{\psi}) \propto  \sigma_1^{-2(a_0-1)} e^{-b_0/\sigma_1^2} \cdot \sigma_2^{-2(a_0-1)} e^{-b_0/\sigma_2^2} \cdot b_0^{\kappa - 1} e^{-b_0/\eta} \\
      \cdot (\left |\bm{\Sigma}_\gamma \right | |\widetilde{\bm{\Sigma}} |)^{-\frac{1}{2}} \exp \left [{\textstyle - \frac{1}{2}} \left\{(\bm{y}-\widetilde{\bm{X}}_{\gamma}\widetilde{\bm{\beta}})' \bm{\Sigma}_{\gamma}^{-1} (\bm{y}-\widetilde{\bm{X}}_{\gamma}\widetilde{\bm{\beta}})  + ( \widetilde{\bm{\beta}} - \widetilde{\bm{\mu}} )' \widetilde{\bm{\Sigma}}^{-1}\widetilde{\bm{\Lambda}} ( \widetilde{\bm{\beta}} - \widetilde{\bm{\mu}} ) \right \} \right] \text{,}
\end{multline}
where $\widetilde{\bm{X}}_{\gamma} = \bm{X}_{\gamma}\bm{T}_\gamma\in \mathbb{R}^{n \times p}$ and, in analogy to the quantities $(\bm{\Sigma},\bm{\mu},\bm{\Lambda})$, we define $\widetilde{\bm{\Sigma}} \triangleq \operatorname{diag}(\sigma_1^2 \bm{1}_{d_1+1}, \sigma_2^2 \bm{1}_{d_2}) \in \mathbb{R}^{p \times p}$, $\widetilde{\bm{\mu}} = (\bm{\mu}_1', \widetilde{\bm{\mu}}_2')'\in\mathbb{R}^{p \times 1}$, and $\widetilde{\bm{\Lambda}} = \operatorname{diag}(\bm{\Lambda}_1, \widetilde{\bm{\Lambda}}_2)\in\mathbb{R}^{p \times p}$.

The transformation $\bm{T}_\gamma$ makes it possible to integrate out the regression coefficients $\widetilde{\bm{\beta}}$ using standard manipulations. Indeed, introducing the terms $\tilde{\bm{A}}_\gamma \triangleq  {\widetilde{\bm{X}}_{\gamma}}' \bm{\Sigma}_\gamma^{-1}\widetilde{\bm{X}}_{\gamma} + \widetilde{\bm{\Sigma}}^{-1}\widetilde{\bm{\Lambda}}\in \mathbb{R}^{p \times p}$ and $\tilde{\bm{b}}_\gamma \triangleq \widetilde{\bm{\Lambda}}\widetilde{\bm{\Sigma}}^{-1}\widetilde{\bm{\mu}} + {\widetilde{\bm{X}}_{\gamma}}'\bm{\Sigma}_\gamma^{-1}\bm{y}\in \mathbb{R}^{p \times 1}$ as before, we obtain the marginal posterior
\begin{multline}
        \label{eq:RB1ContPosterior}
         p(\gamma, \sigma_1^2, \sigma_2^2, b_0 | \bm{y}; \bm{\psi}) \propto \sigma_1^{-2(a_0-1)} e^{-b_0/\sigma_1^2} \cdot \sigma_2^{-2(a_0-1)} e^{-b_0/\sigma_2^2} \cdot b_0^{\kappa - 1} e^{-b_0/\eta}\\
         \cdot \left(\left|\bm{\Sigma}_{\gamma} \right |  |\widetilde{\bm{\Sigma}} | \left |\bm{A}_{\gamma} \right |\right)^{-\frac{1}{2}} \exp \left \{{\textstyle - \frac{1}{2}}(\bm{y}'\bm{\Sigma}_{\gamma}^{-1}\bm{y} + \widetilde{\bm{\mu}}'\widetilde{\bm{\Sigma}}^{-1}\widetilde{\bm{\mu}} - {\widetilde{\bm{b}}_{\gamma}}'
         \widetilde{\bm{A}}_{\gamma}^{-1}\widetilde{\bm{b}}_{\gamma} ) \right \} \text{.}
\end{multline}
It is straightforward to generalize this notion to any $s\in\{-1,0,\ldots,d_1+d_2\}$; a prior is put on $d_1 + d_2 - s + 1$ regression coefficients, and a transformation $\bm{T}_\gamma$ analogous to~\eqref{eq:Transform} defined.

As noted previously, the smoothness constraint of~\eqref{eq:contConstraint} introduces dependence among the pre- and post-changepoint regression coefficients.  In contrast to the marginal posterior distribution of~\eqref{eq:RB1Posterior} derived earlier for the unconstrained case, enforcement of~\eqref{eq:contConstraint} precludes integrating out the associated noise variances $\sigma_1^2$ and $\sigma_2^2$.  In the former case, the block-diagonal structure of $\bm{X}_{\gamma}$ (and therefore of $\bm{A}_{\gamma}$) implies that the induced joint distribution on the variances is separable. However, in the latter case of~\eqref{eq:RB1ContPosterior}, $\widetilde{\bm{X}}_{\gamma}$ is \emph{not} block diagonal---owing to the action of $\bm{T}_\gamma$---and hence neither is $\widetilde{\bm{A}}_{\gamma}$. Therefore, the variances $\sigma_1^2$ and $\sigma_2^2$ are no longer conditionally independent, and their joint distribution does not take the form of known generalizations of the univariate Gamma distribution to the bivariate case~\citep{Yue01}.  Consequently, only the conditional distribution of the hyperparameter $b_0$ is in standard form, and simulation from~\eqref{eq:RB1ContPosterior} proceeds with all other variables drawn using Metropolis-Hastings (MH) steps, as shown in Algorithm~\ref{alg2} below.
In contrast to the case of Algorithm~\ref{alg1}, where a mixture kernel was employed purely for reasons of computational efficiency, we emphasize here that such a move is in fact \emph{required} to sample from  the full support $(1,n)$ of the changepoint index, otherwise mixing of the underlying chain is poor. As before, the mixture kernel consisted of a local random walk move and an independent global move drawing from a discrete distribution derived as a pointwise evaluation of~\eqref{eq:RB1ContPosterior} on the integers $1,2,\ldots,n$. The coupling of noise variances suggests a joint MH move, but this requires specification of a proposal covariance; in simulations we found separate Normal random walk moves for $\ln(\sigma_1^2)$ and $\ln(\sigma_2^2)$ to be adequate.
\begin{algorithm}
\caption{Smoothness-constrained Gibbs sampler for changepoint estimation} \label{alg2}
\begin{enumerate}
\item Draw $\gamma \sim p(\gamma | \sigma_1^2, \sigma_2^2, b_0, \bm{y}; \bm{\psi})$ according to~\eqref{eq:RB1ContPosterior} using a Metropolis-within-Gibbs step;

\item Draw $\sigma^{2}_1 \sim p(\sigma_{1}^{2}| \gamma, \sigma^{2}_2, b_0, \bm{y}; \bm{\psi})$ likewise;
\item Draw $\sigma^{2}_2 \sim p(\sigma_{2}^{2}| \gamma, \sigma^{2}_1, b_0, \bm{y}; \bm{\psi})$ likewise;

\item Draw $b_0 \sim  p(b_0 | \gamma, \sigma^{2}_1, \sigma^{2}_2, \bm{y};
\bm{\psi}) = \mathcal{G}\!\left(\kappa, \eta^{-1}+\sigma_1^{-2} + \sigma_2^{-2}\right)$.
\end{enumerate}
\vspace{-1.00\baselineskip}%
\end{algorithm}

We re-emphasize that in our experience, integrating out the regression coefficients is essential in order to obtain a Gibbs sampler with favorable mixing properties. In particular, a sampler drawing for each parameter of~\eqref{eq:contPosterior} in the presence of smoothness constraints is severely restricted in its ability to explore the state space if one-coordinate-at-a-time updates are employed. When $s$ continuous derivatives are enforced at the changepoint, the likelihood function becomes more nonlinear in changepoint $x_\gamma$ as $s$ increases. Thus, as the induced constraint set becomes more nonlinear, local moves on the scale of the regression coefficients themselves will lie far from it---leading to small acceptance probabilities. To overcome these problems, one may design a high-dimensional MH move to update all the smoothness-constrained regression coefficients jointly with the changepoint; however, a unique move must be designed for each $s$ to be considered.  In contrast, integrating out the regression coefficients obviates this need by handling such constraints for all $s$ simultaneously.

\vspace{-.2\baselineskip}%
\section{Experimental validation}
\vspace{-.05\baselineskip}%
\label{sec:validation}

In order to experimentally validate Algorithms~\ref{alg1} and~\ref{alg2} prior to their application in the setting of atomic force microscopy, we designed and performed two sets of special macro-scale indentation experiments in which precise contact point identification was made possible by the use of an impedance-measuring electrode mounted on the indenter.  While this direct measurement procedure is precluded in the vast majority of biomaterials experiments involving cells and tissues, as such samples are submerged in an aqueous solution, we were able to measure the true contact point for experiments involving respectively cantilever bending and silicone indentation, as described below.

In addition, we conducted a number of simulation studies to characterize uncertainty in changepoint estimation as a function of various model parameters. On the basis of these simulation studies and subsequent experimental validation, we found both algorithms to be insensitive to the exact choice of hyperparameters $\bm{\psi}$, and hence retained the following settings throughout: $\bm{\mu} = \bm{0}$, $\bm{\Lambda} = 10^{-5}\bm{I}_{p \times p}$, $a_0 = 2,$ $\kappa = 1$, and $\eta = 10^{-2}$, with $p$ being the total number of regression coefficients. We set the pre-contact polynomial degree $d_1 = 1$ throughout, based on the discussion in Section~\ref{sec:data} supporting the assumption of a linear pre-contact regime. When smoothness constraints were used, we retained identical parameter settings, and decremented the value of $p$ appropriately. All posterior distributions were obtained by running the appropriate Gibbs samplers for $50\,000$ iterations, and discarding the first $5\,000$ samples. Convergence was assessed using standard methods confirming that increasing the number of Gibbs iterations did not appreciably change the resulting inference.

\subsection{Validation of changepoint inference via cantilever bending}
\label{sec:cantileverExperiment}

\begin{figure}[!t]
  \centering
  \makebox{\includegraphics[width= .9\columnwidth]{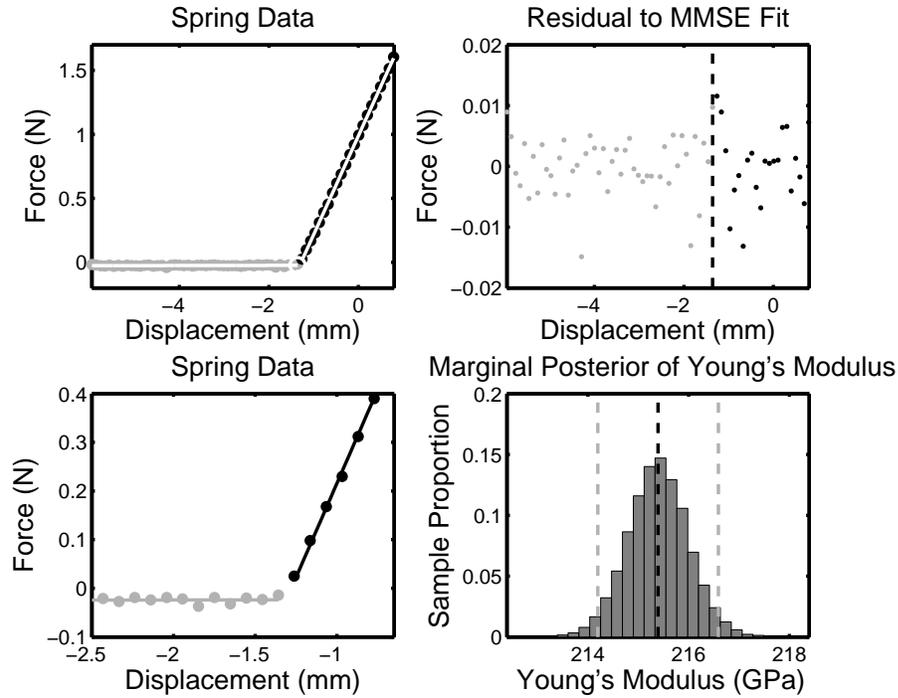}}
  \caption{\label{fig:springPosterior}Inference for the cantilever experiment of Section~\ref{sec:cantileverExperiment}. The force-displacement data and posterior mean reconstructions obtained via Algorithm~\ref{alg1} are shown (top left) with a close-up view near the changepoint (bottom left) and the associated residuals (top right). The marginal posterior of Young's modulus (bottom right) is shown with its mean (black line) and $95\%$ interval (grey lines).}
  \vspace{-.5\baselineskip}%
\end{figure}

We first performed several trials of an experiment whereby a steel cantilever was bent through application of a downward uniaxial force.  Here, the measured force $F$ is expected to change linearly with displacement $\delta = x-x_\gamma$ according to Hooke's Law, with $\gamma \in (1,n)$ representing the contact point index.  Representative data shown in Figure~\ref{fig:springPosterior} were obtained using a TestBench\texttrademark~Series system with a high-fidelity linear actuator (Bose Corporation EnduraTEC Systems Group, Minnetonka, Minnesota, USA), which moved a cylindrical indenter into a cantilevered piece of FSS-05/8-12 spring steel measuring approximately $ 1.27$~cm $\times$ $2.54$~cm (carbon content $0.9$--$1.05$\%; Small Parts, Inc., Miramar, Florida, USA) at a speed of $10$~mm/s.  According to the impedance measurement technique described above, the contact point index $\gamma$ was determined to correspond to position index $48$ of the indenter, corresponding to a contact point $x_\gamma \in [-1.358,-1.262]$~mm.  Because of the hardness of steel and the speed of the indenter, we did not make a smoothness assumption, and thus used the Gibbs sampler of Algorithm~\ref{alg1} to evaluate the efficacy of our approach, with $p=4$ based on the linear post-contact regime implied by Hooke's law.

A full 100\% of posterior values for $\gamma$ after a 10\% burn-in portion took the value $48$, indicating correct detection of the changepoint.  Figure~\ref{fig:springPosterior} shows the results of the curve fitting procedure,with similar results obtained for varying indenter speeds. The minimum mean-square error (MMSE) estimate of Young's modulus was determined to be $215.3$~GPa, with an associated $95\%$ posterior interval of $[214.0,216.6]$.  By comparison, the range of values of Young's modulus for steel with similar carbon content is reported in the literature to be $210 \pm 12.6$~GPa~\citep{Kala05}. Despite its simple design, this experiment is not far removed from practice; a nearly identical procedure was employed by~\citet{wong97} to probe the mechanical properties of silicone-carbide nanorods, with each nanorod pinned on a substrate and subjected to a bending force along the unpinned axis.

\subsection{Analysis of silicone indentation data}
\label{sec:silicone}

While changes in slope at the contact point tend to be more pronounced in harder materials such as steel, the change in measured force is typically smoother in softer materials such as cells and tissues, making the contact point harder to detect.  In earlier work~\citep{YuenRudoy07} we detailed an indentation experiment using a soft silicone sample (Aquaflex; Parker Laboratories, Fairfield, New Jersey, USA), chosen both for its similarity to human tissues used in material indentation studies~\citep{chen1996}, and because it enables direct contact point determination via an impedance-measuring electrode.

\begin{table}
    \caption{\label{tab:silicone} Contact point estimates and associated errors (mm, \%) for the ten silicone trials of Section~\ref{sec:silicone}, based on unconstrained (U) and continuity-constrained (C) models, with force-response data sampled every 0.01~mm on average. Young's modulus estimates $\widehat{E}$ and 95\% posterior intervals for the unconstrained case are also shown, along with averages over all ten trials where appropriate.\vspace{0.5\baselineskip}}
    \centering
    \hspace{-0.04cm}\begin{tabular}{c | r r r r r r r r r r r}
     Trial No. & 1 & 2 &3 & 4 & 5 & 6 & 7 & 8 & 9 & 10 & $\!$Avg.\\
     \hline
     Truth ($x_\gamma$) & $5.52$ & $5.49$ & $5.47$ & $5.44$ & $5.42$ & $5.38$ & $5.39$ & $5.48$ & $5.42$ & $5.38$ & -- \\
     MMSE   (U)         & $5.42$ & $5.44$ & $5.42$ & $5.43$ & $5.47$ & $5.39$ & $5.46$ & $5.35$ & $5.41$ & $5.39$ & -- \\
     MMSE   (C)         & $5.40$ & $5.43$ & $5.41$ & $5.39$ & $5.46$ & $5.37$ & $5.41$ & $5.34$ & $5.34$ & $5.39$ & -- \\
     $\!$Extrap.~(U)$\!$        & $5.39$ & $5.42$ & $5.40$ & $5.37$ & $5.42$ & $5.36$ & $5.40$ & $5.32$ & $5.34$ & $5.38$ & -- \\

     \\
     \%~Err. (U) & $\!\!-1.78$ & $\!\!-0.79$ & $\!\!-0.83$ & $\!\!-0.08$ & $0.88$ & $0.19$  & $1.40$ & $\!\!-2.23$ & $\!\!-0.10$ & $0.11$ & $0.83$ \\
     \%~Err. (C) & $\!\!-2.06$ & $\!\!-1.02$ & $\!\!-1.13$ & $\!\!-0.95$ & $0.31$ & $\!\!-0.25$ & $0.47$ & $\!\!-2.44$ & $\!\!-1.41$ & $0.03$ & $1.00$ \\

     \\
     2.5\%   & $16.5$ & $16.5$ & $16.4$ & $16.4$ & $16.5$ & $16.1$ & $16.6$ & $16.1$ & $16.2$ & $16.4$ & -- \\
     $\widehat{E}$ (kPa) & $17.2$ & $17.3$ & $17.2$ & $17.0$ & $17.2$ & $16.8$ & $17.4$ & $16.8$ & $17.0$ & $16.9$ & $17.1$ \\
     97.5\%  & $17.9$ & $18.4$ & $17.9$ & $17.6$ & $18.0$ & $17.6$ & $18.1$ & $17.7$ & $17.8$ & $17.9$ & -- \\
    \end{tabular}
    \end{table}

Ten trials of this experiment were conducted, using a sample roughly $20$~mm in depth, with a maximum indentation of approximately $8$~mm; a typical force-displacement curve was shown earlier in Figure~\ref{fig:fdExample}. A hemispherical metal indenter of radius $R = 87.5$~mm compressed the sample at $10$~mm/s, and the resulting forces were measured approximately every $10$~$\mu$m to yield $\approxeq 960$ force-displacement data points. Both the unconstrained ($s = -1$) and continuity-constrained ($s=0$) models were fitted in this setting, by way of Algorithms~\ref{alg1} and~\ref{alg2} respectively. For a spherically-tipped indenter, the Hertzian model of~\eqref{eq:indenterEqns} indicates that force is proportional to $(x-x_\gamma)^{3/2}$, and hence we employed a post-contact design matrix with only the fractional power $3/2$. In this regime, we have that $p=4$ for Algorithm~\ref{alg1} and $p=3$ for Algorithm~\ref{alg2}. Since the initial distance from the indenter to the sample was approximately known, a uniform prior on $\gamma \in [125,250]$ was assumed.

For each of the ten data sets collected, the first $n=450$ data points were taken to represent a conservative estimate of operation within the small-deformation regime, and the Gibbs samplers of Algorithms~\ref{alg1} and~\ref{alg2} were each run on these data. Results for both the cases are summarized in Table~\ref{tab:silicone}, along with the experimentally-determined contact point, which varied from trial to trial due to viscoelastic effects. Indeed, given that the spacing between data points is $0.01$~mm on average, it may  be deduced from Table~\ref{tab:silicone} that the average error of $0.8$--$1\%$ across trials corresponds to 8--10 sampled data points.

\begin{figure}[!t]%
    \begin{center}\hspace{-2em}%
    \includegraphics[width=.9\columnwidth]{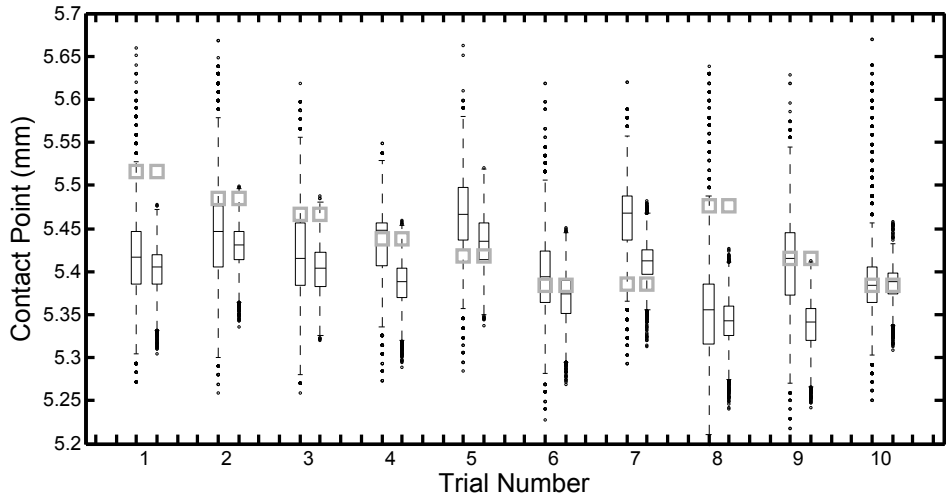}%
    \end{center}
    \vspace{-0.25cm}%
    \caption{\label{fig:siliconeBoxPlots} Box plots of contact point marginal posterior distributions in each of ten silicone indentation trials, shown side-by-side for the unconstrained (left) and continuity-constrained (right) models, with centers of the grey squares indicating true contact point values. Note the consistent decrease in posterior variance and slight downward shift of the posterior under the assumption of continuity.}%
\end{figure}

Marginal contact point posterior distributions for both the unconstrained and continuity-constrained cases are summarized, pairwise by experiment, in the box plots of Figure~\ref{fig:siliconeBoxPlots}.  These  are seen to be notably more diffuse in the former case (left) than the latter (right); this is consistent with the softness of the silicone sample under study, which makes it difficult to reject \emph{a priori} the possibility of continuity at the changepoint.  Absent this assumption, the experimentally determined contact point lies within the $95\%$ posterior interval for each of the ten  trials.  Once a continuity constraint is imposed, the marginal posterior distributions of the changepoints tighten noticeably; this is consistent with our expectation that constraining the model reduces the number of high-likelihood fits.

A more subtle point can also be deduced from the slight yet consistent left-shift across $95\%$ posterior intervals under the continuity-constrained regime relative to the unconstrained case.  Observe the fourth row of Table~\ref{tab:silicone}, which reports the values obtained upon extrapolating pre- and post-contact MMSE curve fits to their meeting point in this latter case.  The increase in force after contact implies that these values will always lie below the directly inferred contact point, as indeed they do.  Nevertheless, enforcing continuity results in only a small increase in overall contact point estimation error, and produces what practitioners might judge to be more physically feasible curve fits.  While independent verification of the Young's modulus is not available for the silicone sample used in our study, the quality of our contact point estimates relative to a known ground truth leads to high confidence in the inferred values of Young's modulus.

Overall, the ability to obtain inferential results and accompanying uncertainty quantification, as exemplified by the cantilever bending and silicone indentation experiments, represents a significant improvement upon current methods; a more in-depth comparison to the method of~\citet{costa} is provided in our earlier work~\citep{YuenRudoy07}.  In particular, the latter experiment demonstrates that reliable estimates of soft material properties can be obtained even in the presence of measurement error---a regime applicable to many AFM  studies of cells and tissues, as we now describe.

\section{Inference in the setting of atomic force microscopy}
\label{sec:indentExp}

Having validated our algorithms on a macroscopic scale, we turn to analyzing cellular biomaterials data collected using atomic force microscopy techniques.  In contrast to our earlier examples, no direct experimental verification is available in this case, though we note that our resultant estimates of Young's modulus are considered plausible by experimentalists (Socrate and Suresh Labs, personal communications, 2008).

It is widely believed that cell biomechanics can shed light on various diseases of import, and hence a key research objective following the advent of AFM technology has been to analyze quantitatively the mechanical properties of various cell types. Indeed, numerous papers over the past decade have linked stiffness and other related mechanical properties to cell malfunction and death~\citep{costaAFM}; for example, the ability of cardiac myocytes in heart muscle tissue to contract is intimately linked to their cytoskeletal structure and its influence on cellular mechanical response~\citep{lieber2004}. Here, we are similarly motivated to understand the stiffness properties of neuronal and red blood cells---currently a topic of intensive research in the biomechanics and bioengineering communities.

\subsection{Indentation study of embryonic mouse cortical neurons}
\label{sec:neurons}

We first analyzed ex-vivo live mouse neurons, submerged in cell culture medium and repeatedly indented by an AFM (Asylum Research, Santa Barbara, California, USA) equipped with a spherically-tipped probe. The indentation of each neuron (Socrate Lab, Massachusetts Institute of Technology) yielded approximately $700$ force measurements, of which the first $500$ were used in subsequent analysis in order to stay within the small-deformation regime.  Such data are of wide interest to neuroscientists and engineers, as traumatic damage to neurons is hypothesized to be related to their mechanical properties.

\begin{figure}[!t]
 \centering
 \makebox{\includegraphics[width= .9\columnwidth]{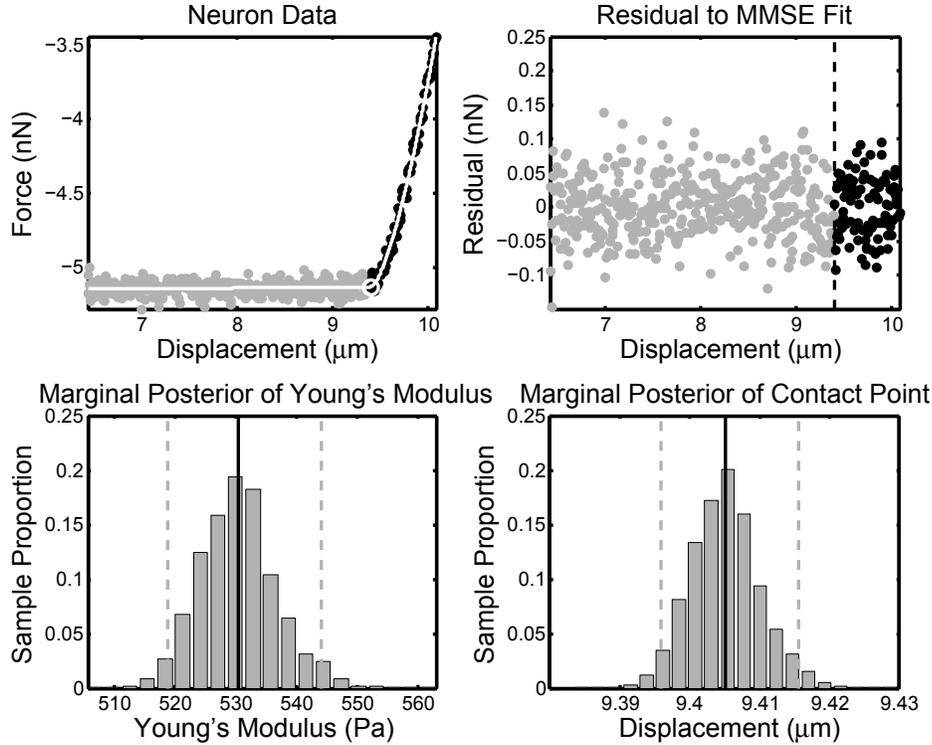}}
  \caption{\label{fig:neuron2Data} Data collected during an AFM indentation of a mouse neuron together with the posterior mean estimate of the underlying regressions (top left) and the induced residuals (top right). Also shown are marginal posterior distributions of Young's modulus (bottom left) and contact point (bottom right), each overlaid with the posterior mean (black line) and 95\% posterior interval (dashed grey lines).}
\end{figure}

As a spherical probe tip was used for indenting each cell, we employed a post-contact design matrix with only the fractional power $3/2$, as in the case of our earlier silicone example, and the continuity-constrained sampler of Algorithm~\ref{alg2}. The results of a typical trial are shown in Figure~\ref{fig:neuron2Data}; the pre- and post-contact residuals were observed to be white.
The primary inferential quantity of interest in such cases is Young's modulus, the principal characterization of cell material stiffness introduced in Section~\ref{sec:ContactPointEst}. According to the Hertzian model of~\eqref{eq:indenterEqns} for a spherical indenter, the regression coefficient corresponding to the $(x-x_\gamma)^{3/2}$ term of the fitted model is proportional to Young's modulus, with the constant of proportionality a function of the given radius $R = 10$~$\mu$m and Poisson's ratio $\nu=0.5$. Thus, we can obtain the posterior distribution of Young's modulus by appropriately scaling the distribution of this regression coefficient, as shown in the bottom left panel of Figure~\ref{fig:neuron2Data}.  We report the MMSE estimate of Young's modulus as $\widehat{E}= 530.4$~Pa, and the corresponding $95\%$ posterior interval as $[518.8,544.1]$~Pa. This estimate is in reasonable agreement with those previously reported for similar neurons~\citep{Lu06,Elkin07}, with variability due primarily  to differences in indentation speeds (Socrate Lab, personal communication, 2008).

\subsection{Indentation study of red blood cells}
\label{sec:rbc}

\begin{figure}[t]
    \centering
    \makebox{\includegraphics[width= .9\columnwidth]{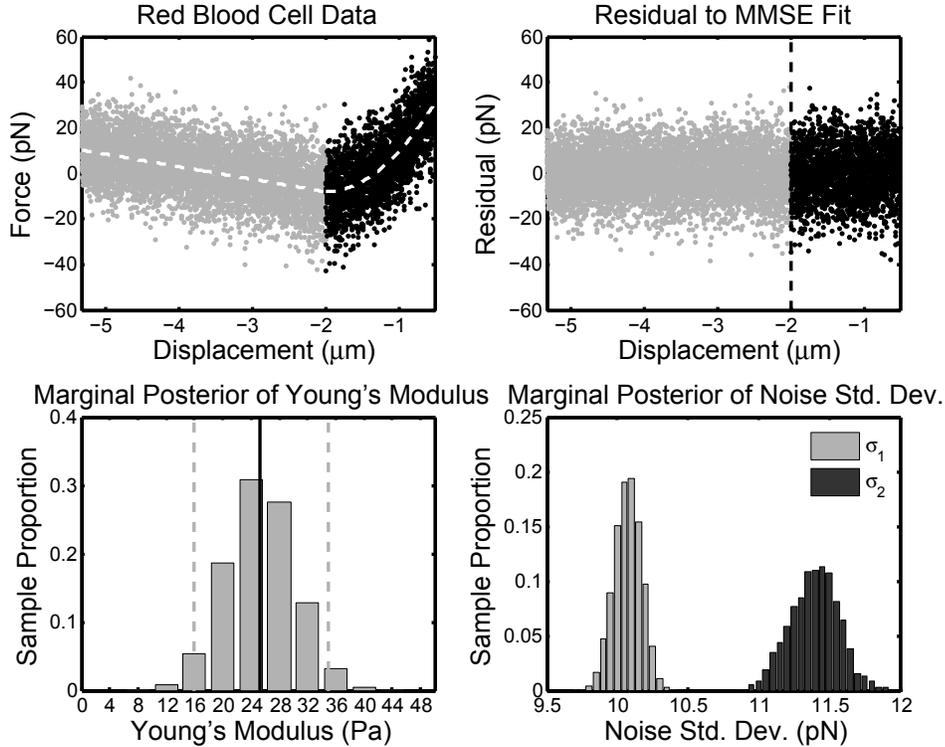}}
\caption{\label{fig:rbc3Data} Data collected during an AFM indentation of a human red blood cell together with the posterior mean estimate of the underlying regressions (top left) and the induced residuals (top right). Also shown are marginal posterior distributions Young's modulus (bottom left) and of $\sigma_1^2$ and $\sigma_2^2$ (bottom right), the former overlaid with the posterior mean and 95\% posterior interval.}
   \vspace{-.5\baselineskip}%
   \end{figure}

The mechanics of red blood cells have also been extensively studied using atomic force microscopy~\citep{radmacher1996}. In this vein, we next analyzed data from ex-vivo live human erythrocytes (red blood cells, Suresh Lab, Massachusetts Institute of Technology) submerged in a cell culture medium, indented by an AFM (Asylum Research) equipped with a pyramidally-tipped probe.  The final $800$ of approximately $8\,500$ data points were discarded prior to analysis, as they clearly lay outside the small-deformation regime. Relative to the neuron AFM data considered in Section~\ref{sec:neurons}, the  sampling rate of resistive force here is high, and consequently it is feasible to enforce continuity ($s=0$) at the changepoint. Algorithm~\ref{alg2} was therefore employed, with
results from a typical trial shown in Figure~\ref{fig:rbc3Data}.

In the case at hand, the regression coefficient $\beta_{12}$ corresponding to $(x-x_\gamma)^{2}$ is proportional to the Young's modulus as per~\eqref{eq:indenterEqns}, with the constant of proportionality depending on the indenter tip angle $2\phi=70^\circ$ and Poisson's ratio $\nu=0.5$, and, as we discuss below, no linear post-contact regression term was included. The inferred distribution of Young's modulus may be obtained by appropriately transforming the marginal posterior of $\beta_{12}$, as detailed in Section~\ref{sec:likelihood}, and is shown in the bottom left panel of Figure~\ref{fig:rbc3Data}. The resultant MMSE estimate of $\widehat{E}=25.3$~Pa and corresponding posterior interval of $[16.0, 34.9]$~Pa were confirmed to be consistent with various experimental assumptions (Suresh Lab, personal communication, 2008). Further, the pre- and post-contact error variances are determined to be unequal, as shown in the bottom right panel of Figure~\ref{fig:rbc3Data}.

Note that in the absence of a post-contact drift, enforcing continuous differentiability at the changepoint ($s = 1$)  constrains the pre-contact linear fit to have zero slope, and is inconsistent with the pre-contact drift clearly visible in the top left panel of Figure~\ref{fig:rbc3Data}. On the other hand, if the post-contact polynomial were to include a linear drift term, then enforcing continuous differentiability would imply that the pre- and post-contact drifts are identical.  Though this is appealing from a modeling viewpoint, as it eliminates an additional free parameter, practitioners lack evidence for such an equality.  Moreover, in our experiments its inclusion had no appreciable effect on the inference of Young's modulus, and so we did not include a post-contact drift term in our final analysis of these data.

As in the case of our earlier experiments, we compared our approach to the likelihood method of~\citet{costa}, which yielded an estimate of $x_\gamma$ shifted to the right by more than $1 000$ data points relative to that shown in Figure~\ref{fig:rbc3Data}.  The corresponding estimate of Young's modulus in turn was found to be $34.8$~Pa---an increase of $37.5$\% relative to the MMSE point estimate of $25.3$~Pa, and close to the upper boundary of our estimated posterior interval.  While in this experimental setting one cannot conclude that either estimate is superior to the other, we note that the difference between them can be attributed in part to our model's incorporation of differing pre- and post-contact error variances, and smoothness constraints.  Thus, one may view our inferential procedures as both a formalization and an extension of earlier likelihood-based approaches developed by practitioners, enabling both robust, automated fitting procedures and explicit uncertainty quantification.

\section{Discussion}
\label{sec:discussion}

In this article we have posed the first rigorous formulation of---and solution to---the key inferential problems arising in a wide variety of material indentation systems and studies. In particular, practitioners in the materials science community to date have lacked accurate, robust, and automated tools for the estimation of mechanical properties of soft materials at either macro- or micro-scales~\citep{Lin2007a, crick07}. A principal strength of our approach is its applicability to the analysis of biomaterials data obtained by indenting cells and tissues using atomic force microscopy; contact point determination is even more difficult in this setting, owing to the gradual change of measured resistive force that is a hallmark of soft materials. The Bayesian switching polynomial regression model and associated inferential procedures we have proposed provide a means both to determine the point at which the indenting probe comes into contact with the sample, and to estimate the corresponding material properties such as Young's modulus. In turn, its careful characterization holds open the eventual possibility of new biomechanical testing procedures for disease~\citep{costaAFM}.

Our parametric approach is strongly motivated by---and exploits to full advantage---the Hertzian models governing the physical behavior of linear-elastic materials undergoing small deformations. The Bayesian paradigm not only enables uncertainty quantification, crucial in applications, but also allows for the natural incorporation of physically-motivated smoothness constraints at the changepoint. Inference is realized through application of carefully designed Markov chain Monte Carlo methods together with classical variance reduction techniques. The resultant algorithms have been shown here to be both statistically and computationally efficient as well as robust to choice of hyperparameters over a wide range of examples, and are available online for use by practitioners. Indeed, the direct applicability of our methods precludes any need for data pre-processing prior to analysis.

Outside of the linear-elastic materials we consider here, it is of interest to apply the methodology to viscoelastic materials (e.g., biopolymers) which return to their pre-contact state slowly over time~\citep{Lin2007b}. In such cases, the amount of induced deformation depends not only on the indenter geometry, but also on the rate of indentation. Another extension is to incorporate multiple spatially distributed changepoints into our model, a key construct when atomic force microscopes are used to indent repeatedly a sample in order to characterize cell stiffness as a function of surface location~\citep{Geisse09}. Finally, a sequential estimation scheme could be of great use in surgical robotics applications, where contact point determination plays a key role in enabling tactile sensing---a subject of current study by the authors.

\section*{Acknowledgements}

The authors would like to acknowledge Kristin Bernick, Anthony Gamst, Hedde van Hoorn, Petr Jordan, and Thibault Prevost for helpful discussions, and would especially like to thank Simona Socrate and Subra Suresh for providing access to data from atomic force microscope indentation experiments on neurons and red blood cells, respectively. The first author is sponsored by the National Defense Science and Engineering Graduate Fellowship. The second author is sponsored by United States National Institutes of Health Grant No.~NIH R01 HL073647-01.  The authors are grateful to anonymous reviewers for many suggestions that have helped to improve the clarity of this article.

{\small
\begin{spacing}{0.5}

\end{spacing}}

\end{document}